\documentclass[sigconf,nonacm]{acmart}
\usepackage{caption}
\usepackage{multirow}
\usepackage{xcolor}
\usepackage{graphicx}
\usepackage{epstopdf}
\usepackage{graphics}
\usepackage{url}
\usepackage{comment}

\AtBeginDocument{%
  \providecommand\BibTeX{{%
    \normalfont B\kern-0.5em{\scshape i\kern-0.25em b}\kern-0.8em\TeX}}
    }

\begin{document}

\title{Company Similarity using Large Language Models}



    
\author{Dimitrios Vamvourellis}
\email{dimitrios.vamvourellis@blackrock.com}
\affiliation{%
  \institution{BlackRock, Inc.}
  \city{New York}
  \state{NY}
  \country{USA}
}
\author{M\'{a}t\'{e} T\'{o}th}
\email{mate.toth@blackrock.com}
\affiliation{%
  \institution{BlackRock, Inc.}
  \city{Budapest}
  \country{Hungary}
}
\author{Snigdha Bhagat}
\email{snigdha.bhagat@blackrock.com}
\affiliation{%
  \institution{BlackRock, Inc.}
  \city{Gurugram}
  \state{Haryana}
  \country{India}
}
\author{Dhruv Desai}
\email{dhruv.desai1@blackrock.com}
\affiliation{%
  \institution{BlackRock, Inc.}
  \city{New York}
  \state{NY}
  \country{USA}
}
\author{Dhagash Mehta}
\email{dhagash.mehta@blackrock.com}
\affiliation{%
  \institution{BlackRock, Inc.}
  \city{New York, NY}
  \country{USA}
}
\author{Stefano Pasquali}
\email{stefano.pasquali@blackrock.com}
\affiliation{%
  \institution{BlackRock, Inc.}
  \city{New York, NY}
  \country{USA}
  }

\renewcommand{\shortauthors}{Vamvourellis et al.}

\begin{abstract}
Identifying companies with similar profiles is a core task in finance with a wide range of applications in portfolio construction, asset pricing and risk attribution. When a rigorous definition of similarity is lacking, financial analysts usually resort to 'traditional' industry classifications such as Global Industry Classification System (GICS) which assign a unique category to each company at different levels of granularity. Due to their discrete nature, though, GICS classifications do not allow for ranking companies in terms of similarity. In this paper, we explore the ability of pre-trained and finetuned large language models (LLMs) to learn company embeddings based on the business descriptions reported in SEC filings. We show that we can reproduce GICS classifications using the embeddings as features. We also benchmark these embeddings on various machine learning and financial metrics and conclude that the companies that are similar according to the embeddings are also similar in terms of financial performance metrics including return correlation. 

\end{abstract}

\maketitle

\section{Introduction}
One of the crucial tasks for the financial analyst is identifying similar companies, which allows them to compare and benchmark different firms on equal grounds; identify a control group to investigate the effects of some financial or policy intervention; identify outliers with respect to other companies in a sector; compute different risk factors; construct diversified portfolios; discover fair price of a security through relative valuation with respect to its peers; etc. Because of such high importance of identifying company peers, various industry classification systems such as Global Industry Classification System (GICS) \cite{msci2020global}, Standard Industry Classification (SIC) \cite{fama1997industry}, North American Industry Classification System (NAICS) \cite{murphy1998introducing}, etc. are proposed by different research groups and financial companies.

In this work, we focus on GICS classification which was developed by Morgan Stanley Capital International (MSCI) and Standard \& Poor's (S\& P), and is a popular classification system that includes U.S. and global companies. GICS classification system assigns a unique category to each company at 4 different levels of granularity according to the company's principal business activity (Sector, Industry Group, Industry and Sub-industry from least to most granular). 

However, current industry classifications have several limitations: they fail to capture that many companies operate in multiple industry sectors which might not even be related. Assigning unique pre-defined classification codes to such a company may mislead the analyst to compare the company to a narrow group of peers, missing out on many other aspects of the company's business. In addition, often, the definitions of industry sectors are  fluid, e.g., retail may also include online shopping and brick-and-mortar stores. Moreover, the classification system can only provide qualitative peer group for a target company, but it does not provide a rank ordered list of the peer companies based on a quantitative measure of similarity. In addition, different data vendors may assign different codes to the same businesses \cite{guenther1994differences}.

Company similarity has been extensively investigated using various types of datasets in the literature. We briefly review the existing literature by broadly classifying the literature in terms of the dataset used to investigate company similarity: structured, textual, mixture of structured and textual datasets. 

Historically, structured data has been most extensively used to identify peer groups for companies due to the easier accessibility of such data (the reader is referred to Ref.\cite{damodaran2012investment} for a recent exhaustive review on this topic). Here, typically, firms' fundamentals and accounts-based information are used to identify company peers using traditional statistical methods such as least squared regressions \cite{bhojraj2002my,liu2002equity,rhodes2005valuation,bartram2018agnostic,kaustia2015social}. Other approaches are based on different datasets such as internet co-searches of firms on the SEC website by users \cite{lee2015search}, inter-firm transaction data \cite{dalziel2018can}, patent citations \cite{gay2014patent}, co-mentions of firms in tweets \cite{sprenger2011tweets}, etc. Recently, machine learning methods have also been applied on such structured data to identify company peers \cite{geertsema2023relative}. 

Other structured data that is usually exploited to directly investigate the company peers, and, more specifically, industry classifications, are stock returns correlations. In Ref.~\cite{chan2007industry} (see also Ref.~\cite{jung2016clustering}), mean correlations within individual GICS class at various levels of hierarchy were computed to show that common movement in returns that may be attributed to the companies being in the same industry class is stronger for stocks of large companies than for those of small companies. Various Refs.~\cite{saha2022survey,saha2021stock,sarmah2022learning} have employed network science and graph ML based methods to learn lower dimensional embeddings of stock returns of different companies. Refs.~\cite{dolphin2022stock,dolphin2023stock, dolphin2023industry,long2020integrated,yi2022stock2vec} employed natural language processing inspired methods to learn the embeddings using the stock returns data. In all these works, eventually company similarities are computed within the respective embeddings and then compared against GICS classification. Note that these works attempt to learn embeddings in an unsupervised manner, i.e., the GICS classes are not supplied into the algorithm as the target variable.

Also, another category of approaches is based on the the use of text data to learn company similarity using Natural Language Processing (NLP) techniques. In the financial field, various NLP techniques including older dictionary-based approaches, have been widely used to extract information from textual data such as SEC filings, news articles and social media posts ~\cite{loughran2020textual,hoberg2016text}. In Ref. \cite{hoberg2016text}, the authors proposed a text-based classification system based on business descriptions from 10K filings and demonstrated that their classification outperformed SIC and NAICS in terms of similar profitability, sales growth, and market risk. In Ref.~\cite{lu2021stock}, company embeddings were computed using stock news and sentiment dictionaries to predict stock trends. In Ref.~\cite{wu2019deep} company embeddings were computed based on the co-occurrence matrix of stocks obtained by counting the number of times pairs of companies were mentioned in news articles using Glove algorithm \cite{pennington2014glove}. 

In Ref.~\cite{rizinski2023company}, the authors posed the problem of classifying companies as a zero-shot learning problem where a pre-trained transformers based model was fed with the company descriptions available in the Wharton Research Data Services (WRDS) dataset without any finetuning on the model to classify the companies to their GICS classes. Though the eventual goal of this work is to potentially automate the process of GICS classification, the accuracy metrics reported is in the range of 0.64 (weighted average $F_1$ score). In Ref.~\cite{ito2020learning}, authors used SEC 10K for the US companies and similar forms for the Japanese companies to finetuned BERT (originally trained separately on the English and Japanese corpora) on three tasks: predicting the sector labels, capturing stock market performance and modeling sector names. 

Finally, some recent works have attempted to combine various types of datasets and used ML techniques to learn similarity among companies to compare them with GICS classifications. Most notably, in Ref. ~\cite{bonne2022artificial} the authors use features based on tabular data such as returns and factor exposures, Node2Vec embeddings based on news co-mentions as well as Term Frequency- Inverse Document Frequency (TF-IDF) and Doc2Vec embeddings of 10K filings. Given these features, they used cosine and ML distance metrics to calculate the similarity scores between pairs of companies on year $t$. These pairwise-similarity scores are used as features in ML models (ridge regression, neural networks and XGBoost) to predict the similarity between companies on year $t+1$, as this is measured by the correlation of the daily returns between two companies in a given pair.




\subsection{Our Approach and Contributions}
In the present work, we do not intend to construct yet another industry classification system. Instead, our goal is to construct company embeddings based on the business description reported in SEC 10K filings and explore their performance on multiple downstream financial tasks. In particular, the goals of this paper are: (1) to explore if the GICS classifications can be reproduced from the original using state-of-the-art (SOTA) NLP techniques; (2) to benchmark company embeddings generated from different language models on various financial downstream tasks; and (3) study the effects of different factors (such as the pre-training objective, effect of finetuning, model size) on the quality of text embeddings generated by various SOTA language models as measured by various ML and financial metrics.

The remainder of the paper is organized as follows: In section \ref{sec:data_desc} we provide a description the data used and preprocessing steps taken. In section \ref{sec:methodology}, we outline the methodology used to conduct our experiments. In section \ref{sec:results}, we discuss the results of the experiments conducted, in section \ref{sec:usecases} we present sample financial applications of the company embeddings followed by the conclusions drawn in section \ref{sec:conclusion}.

\section{Data Description and Preprocessing}\label{sec:data_desc}

\subsection{Data Description}\label{sec:data}
In this work, we used information from the U.S. Securities and Exchange Commission (SEC) filings which are submitted by public companies and issuers of securities to maintain information symmetry between the stakeholders and investor communities. These filings are publicly available to ensure transparency with respect to company strategy, financial achievements, legal proceedings and management discussions. In particular, we used the textual content from Item 1 of the SEC 10K reports as input features. The 10K reports are issued annually and Item 1 (called the Business Description section) of the form contains information regarding details on the principal products and securities offered, competitive factors, distribution method, markets of operation, etc. Item 1 on average ranges between 1500-1800 words. Here, we consider 2590 10K filings for the year 2022 from the Russell 3000 security universe which contains the top 3000 US companies by market capitalization. The security universe is primarily composed of firms from the Financial (18\%), Health Care (17\%), Information Technology (15\%) and Consumer Discretionary (13\%) sectors. As for the target variable for some of our experiments, we used Sector and Industry levels from GICS hierarchical industry classification system.


\subsection{Data Preprocessing}
We apply standard NLP preprocessing on Item 1 of 10K filings, i.e., removing any Uniform Resource Locators (URLs) and non-ascii characters as well as white spaces. We also transform all the text to lower case. Then, we use the standard BERT-base-uncased tokenizer which tokenizes the input text using WordPiece and a vocabulary size of 30,000 tokens. 


\section{Methodology}\label{sec:methodology}
The main goal of this paper is to study company similarity based on the information reported in the business description section of the annual 10K filings. In order to do so, we need to construct a numerical representation of each company based on the business description. Thus, the underlying method used to generate numerical representations from text is a key consideration. Older approaches focused on representing documents as high-dimensional sparse vectors with length equal to the number of words in the vocabulary and each entry representing the frequency of a given word in the input text. More recently, distributed text representations -- often called embeddings -- were introduced which encode words and in turn entire documents as dense vectors, capturing semantics more effectively. In the past decade, text representation learning has gone through multiple breakthroughs - from static embeddings such as Word2Vec \cite{mikolov2013} to context-aware word embeddings like BERT \cite{devlin2018bert}, sentence embeddings like SBERT \cite{reimers2019sentence} and, more recently, to embeddings extracted from LLMs. Given the importance of the underlying text representation method, we have explored several models to extract text embeddings as detailed in \ref{subsec:models}.

Subsequently, we measure the ability of each model in generating financially meaningful embeddings on multiple downstream tasks using standard ML and financial metrics. Details for each downstream task are presented in section \ref{sec:downstream_task}.

\subsection{Pre-trained vs Finetuned Models}
In this paper, we focus on extracting embeddings using SOTA pretrained as well as finetuned LLMs. Such models have been trained on vast amounts of data and learned to generate meaningful numerical representations of text which capture the semantic and syntactic structure of language. Models which are pre-trained on a diverse range of general-purpose text may not completely align with this specific domain of data or the intricacies of financial language. To this end, we also explore whether models  finetuned with sample 10K-filed business descriptions and GICS industry labels result in more effective embeddings for downstream tasks. In section \ref{subsec:models}, we outline the steps taken to finetune BERT, SBERT and Longformer respectively.

\subsection{Context Window}
In language modeling terminology, context window refers to the maximum number of tokens which can be fed at a time to a given model to generate a document embedding and in turn perform a task like text classification, text generation or question answering. Different language models use different network architectures and in turn the length of the maximum context window varies depending on the model. For example, BERT and SBERT have a maximum context window size of 512 tokens, Longformer \cite{beltagy2020longformer} supports up to 4096 tokens, GPT-based \cite{gpt} text-embedding-ada-002 supports up 8192 tokens and PaLM-based \cite{palm} text-embedding-gecko@001 supports up to 3072 tokens.

In this study, we also explore whether more information from the business description section leads to semantically richer embeddings for the tasks at hand. Specifically, we benchmark the quality of embeddings generated by each model for the first 512, 1024 and 1536 tokens of the business description. To generate embeddings for documents exceeding the model's maximum context window size, we split the input document in chunks of maximum context size, generate embeddings for each chunk and then average the chunk embeddings to obtain the final document representation. 

\subsection{Models}\label{subsec:models}
Below, we describe the different models used in our experiments, and the finetuning process where applicable.

\subsubsection{BERT-finetuned}
BERT \cite{devlin2018bert} (Bidirectional Encoder Representations from Transformers) is a language model based on the popularized Transformer \cite{vaswani2017attention} architecture. The model is trained in a self-supervised fashion, on the tasks of a) masked language modeling - model is trained to predict randomly masked words in a sentence given right and left context; and, b) next sentence prediction - model is fed with two sentences and is trained to predict whether the second sentence follows the first. 

In this work, we use BERT-base-uncased model as the base model and we finetune it on the task of predicting GICS industry, using only the first 512 tokens of Item 1. Specifically, we stack a softmax layer of 66 dimensions (equal to the number of distinct GICS industries) on top of the pretrained BERT-base model. To prevent "catastrophic forgetting" \cite{kirkpatrick2017overcoming}, we apply gradual unfreezing during finetuning. We first freeze all layers apart from the softmax layer which we train for 15 epochs using a high learning rate of 0.01, mini-batch size of 16 and maximum sequence length of 512 tokens. We then unfreeze all layers and further train the entire model for another 5 epochs using a smaller learning rate of $2e-5$ to prevent the base layers from forgetting basic language information while focusing on this classification task. 

When BERT is used for classification tasks, the [CLS] token is prepended to the input text sequence before feeding it to the model. This special token acts as a summary representation of input text allowing the model to capture the overall semantics needed to perform a downstream task like classification. Here, after finetuning BERT on predicting GICS industries, we use the representation of the [CLS] token from the last encoder layer as the embedding of the input text provided. This results in generating company embeddings of 768 dimensions.

\subsubsection{SBERT} Sentence-BERT \cite{reimers2019sentence} leverages the Transformer architecture and employs siamese and triplet networks to learn effective sentence embeddings. SBERT uses pre-trained BERT as a base and is finetuned on the task of semantic similarity and paraphrase detection. Particularly, SBERT is trained to minimize a task-specific loss like contrastive or triplet loss which encourages sentences with similar meaning to have closer representations while dissimilar sentences are pushed further apart in the embedding space.

We study the quality of embeddings generated from a pretrained general-purpose SBERT model. Specifically, we use all-mpnet-base-v2 model version of SBERT which uses the pre-trained microsoft-mpnet-base \cite{mpnet} model as a base and was finetuned on a 1 billion sentence pairs on a contrastive learning objective.

We also explore the quality of finetuned SBERT embeddings. To finetune SBERT, we create pairs of business descriptions (using only the first 512 tokens of each company description) and we label each pair with 1 if the descriptions belong to companies in the same GICS industry and 0 otherwise. For each company description in the original dataset, we create one positive pair with a randomly chosen company description from the same industry and one negative pair with a description of a company randomly chosen from any other industry. This results in a balanced training dataset of 5180 document pairs. We then finetune SBERT all-mpnet-base-v2 model using cosine similarity loss on this dataset for 3 epochs using a batch size equal to 4 and warmup steps equal to 100. In this way, the model is penalized if it places two business descriptions from the same industry further apart and is encouraged to learn semantically similar representations for companies from the same GICS industry. This  model generates document embeddings of 768 dimensions.

\subsubsection{Longformer-finetuned.} Longformer \cite{beltagy2020longformer} leverages the transformer architecture and employs an attention mechanism that scales linearly in accordance with the sentence length, thus providing an ability to process relatively longer documents and setting new standards on long document classification tasks. This is achieved by combining the local windowed attention that builds local contextual representation along with a task specific global attention that builds full sequence representation for long range contextual information. Longformer uses pretrained RoBERTa \cite{liu2019roberta} as a base and is finetuned for multiple downstream tasks like QA, classification and coreference resolution. 
In this paper, we also experiment with generating embeddings from a LongFormer version finetuned in predicting the GICS Industry classification, using the first 512 or 1024 tokens of Item 1 as input text. For finetuning we apply gradual unfreezing. We first train the classifier layer for 15 epochs with a high learning rate of 0.01 and mini-batch size of 4, then we unfreeze and train the entire model for 5 more epochs with a relatively smaller learning rate of $2e - 5$ and a batch size of 4 so as to make sure that the base layers can process the basic language information along with carrying out the classification task. This  model generates document embeddings of 768 dimensions.

\subsubsection{GPT-based embeddings} Generative Pre-Trained Transformer \cite{gpt,gpt2} is a language model released by OpenAI bassed on the decoder Transformer architecture. It has been trained on multiple data sources like Common crawl dataset (around 600 billion words of text), GitHub dataset (100 million code repository), Stack overflow dataset (170 million questions and answers) on the task of next word prediction and was finetuned on instruction datasets. GPT-based models ($\sim 175$ billion parameters) recently set new state-of-the-art standards across a wide range of natural language understanding tasks including translation, summarization and question answering based on multi-hop reasoning. During these massive pretraining and finetuning phases, the model has learned to generate high-quality compressed representations of textual data. In this study, we explore the ability of the GPT-based text-embedding-ada-002 model to generate embeddings for long documents. This model, which generates embeddings of 1536 dimensions, was published by OpenAI in December 2022 outperforming older OpenAI embedding models on text search, code search, and sentence similarity tasks.

\subsubsection{PaLM-based embeddings} PaLM \cite{palm} (Pathways Language Model) is a dense decoder only Transformer model with 540-billion parameters trained on around 780 billion tokens of text. The primary advantage of PaLM lies in its training that leverages Pathways \cite{barham2022pathways} to enable efficient training of large neural networks across huge number of TPU pods. It has been tested on multiple downstream tasks like Language Modelling, QA, Few-shot learning, Cross Lingual QA, Natural Language Inference, Arithmetic Reasoning etc. It has been trained on multilingual datasets that includes web documents, GitHub code, wikipedia, conversations etc. In this study we are testing the PaLM-based text-embedding-gecko@001 model which generates document embeddings of 768 dimensions.

\subsection{Computational Setup}
All of the experiments outlined above were conducted on a n1-standard-32 GCP instance with 4 TESLA T4 GPUs, 32 CPUs and 120GB of host memory. The 20-epoch finetuning routines outlined above were completed within approximately (wall-clock time): 16 minutes for BERT, 36 minutes for SBERT, 40 minutes for Longformer finetuned on first 512 tokens and 60 minutes for Longformer finetuned on first 1024 tokens.

\subsection{Evaluation Metrics}
In order to evaluate different downstream tasks highlighted in Section \ref{sec:downstream_task} we use an appropriate evaluation metric based on the task. For classification tasks we rely on accuracy as well as micro and weighted F1 score. For validating similarities based on learned embeddings we rely on pairwise correlations with respect to returns. In case of return attribution with respect to different clustering algorithms we rely on explained $R^2$.

\subsection{Downstream Tasks}\label{sec:downstream_task}
\subsubsection{GICS Sector/Industry Classification}\label{sec:sector_industry_classification} Our first approach to  evaluate the effectiveness of the proposed embeddings is by assessing their ability to accurately reproduce GICS classifications. To accomplish this, we utilize the vectors generated by each embedding model as input features to a multinomial logistic regression classifier. In two separate experiments  we use both the GICS Sector and Industry categories as target variables respectively. We use ‘L2’ regularization to improve the robustness of the logistic regression classifier and prevent over-fitting. We used an 80-20 train-test split to examine the out-of-sample performance of each model. Due to the observed imbalance within our GICS classifications, we opt for a stratified split, ensuring a representative sample from all categories in both train and test splits.

\subsubsection{Similarity} We benchmark the above embedding models on the task of identifying the top $k$ peers of each company - well performing language models would place similar companies close to each other in embedding space while dissimilar ones would be placed further apart. In this study, we measure similarity between two companies based on the correlation of their daily returns, assuming that similar companies typically experience similar movements in their stock returns over long horizons. 
For two stocks $i$ and $j$ at the end of year $t$ we calculate $\rho_{ij}$, the pairwise correlation between their daily returns. To evaluate the model's ability to identify similar companies to company $i$, we first calculate the average pairwise correlation with the returns of the closest $k$ neighbors given by $\bar{\rho_i} = \sum_{j=1, i\neq j}^{k} \frac{\rho_{ij}}{k}$. The top $k$ peers are identified based on cosine similarity between the company embeddings. 
After obtaining the average pairwise correlation for each company, we calculate the final metric by averaging over all the companies in the defined universe, $\bar{\rho} = \sum_{i=1}^{K} \frac{\bar{\rho_{i}}}{K}$. We repeat this experiment for years between 2019-2022 and we report the average performance for different values of $k$ in table \ref{table:corr}. For benchmarking purposes, we also calculate the same metric for GICS sector and industry classification by setting $k$ equal to to the number of companies in the same sector/industry in absence of a continuous distance metric (i.e. referring to as dynamic $k$ in table \ref{table:corr}).


\subsubsection{Return Attribution} \label{sec:return_atrib}
We expect that similar companies will react similarly to systematic market risk factors. Hence, identifying clusters of similar companies can be used as risk factors which can explain part of the return which is caused due to market shocks which are common to each cluster. To test whether the embeddings result in financially coherent and meaningful clusters, we test if they can be used to explain historical equity returns. To do this, we first apply different clustering techniques on the generated embeddings and use the cluster assignment as a categorical feature to explain monthly equity returns.

In order to generate optimal clusters we have experimented with several clustering algorithms like kmeans, agglomerative, feature agglomerative and spectral clustering. We first perform reduction of embeddings using UMAP \cite{mcinnes2018umap} followed by generating clusters via all the above listed clustering methods. The input dimensions of embedding generated from the models that have been experimented in this paper have high amount of variability. Thus the optimal number of dimensions to which the embeddings can be reduced was determined based on downstream task performance.
The efficacy of the clusters has been measured using three major clustering evaluation metrics namely Homogeneity, Completeness and V-measure. These metrics serve as an intuitive way of verifying the clustering algorithms. Based on the above metrics it was identified that the spectral clustering consistently had better scores and we use this method to perform clustering in the embedding space. This can be attributed to the fact that spectral clustering performs relatively better in case of higher dimensional data and is more robust to noise and outliers. 

For this experiment, we calculate the cumulative monthly price return for each asset in the universe defined in section \ref{sec:data}. To remove any auto-correlation effect, we then run a separate cross-sectional regression for each month $t$:
$$R_{j,t} = A(t) + \sum_{i=1}^{N} B_{i,t} C_{j,i} + \epsilon_{j,t}$$
where $R_{j,t}$ is the cumulative return of stock $j$ in month $t$, $C_{j,i}$ is an indicator variable which is equal to 1 if stock j belongs to cluster $i$ and $\epsilon_{j,t}$ is the residual return of stock $j$ in month $t$. $A(t)$ is the intercept and $B_{i,t}$ is the return of cluster $i$ in month $t$, both learned by the regression model. For each monthly regression fit we record the $R^2$. We calculate the final metric by averaging over the $R^2$ obtained from monthly regressions fit between January 2019-May 2023 - this is measure of the returns variance explained on average by the clustered company embeddings. Results are reported in table \ref{table:corr}.

\begin{table}[htbp]
 \centering
 \small
 \begin{tabular}{l l l l l l}
 \toprule Method  & \multicolumn{4}{c}{Avg Pairwise Correlation} & $R^2$ \\
 \midrule
 {} & $k=dynamic$ & $k=1$ & $k=5$ & $k=10$ & {}  \\
 \textbf{GICS Sector} & 0.362 & - & - & - & 0.052 \\
 \textbf{GICS Industry} & 0.409 & - & - & - & 0.106 \\
 \textbf{BERT-FT} & - & 0.450 & 0.430 & 0.421 & 0.100  \\
 \textbf{LF-FT-512} & - & 0.449 & 0.425 & 0.415 & 0.110  \\
  \textbf{LF-FT-1024} & - & 0.319 & 0.312 & 0.309 & 0.096  \\
  \textbf{SBERT-FT} & - & 0.458 & 0.438 & 0.428 & 0.100 \\
  \textbf{SBERT-PT} & - & \textbf{0.471} & \textbf{0.443} & \textbf{0.432} & 0.114  \\
  \textbf{GPT-ada} & - & 0.462 & 0.438 & 0.425 & 0.117  \\
  \textbf{PaLM-gecko} & - & \textbf{0.471} & 0.442 & 0.431 & \textbf{0.119} \\
\bottomrule
\end{tabular}
\caption{Performance metrics for similarity (avg pairwise correlation) and return attribution ($R^2$) tasks. Results shown are based on embeddings generated using the first 1536 tokens of the business description.}
\label{table:corr}
\vspace{-6mm}
\end{table}

\section{Results and Discussion}\label{sec:results}

\textbf{Company embeddings can reproduce GICS sector/industry classifications.} Table \ref{table:classification_sector_industry} summarizes the performance of different language models in generating embeddings which can be used as features to directly predict the GICS sector label. Finetuned Sentence-BERT (SBERT-FT) models preform best, achieving approximately 90\% accuracy in predicting GICS sectors from 10-K business descriptions. SBERT without finetuning ranks behind SBERT-FT with an accuracy of 83.6\%. The fact that SBERT models perform well both with and without finetuning can be attributed to the fact that SBERT is designed to produce semantically rich sentence embeddings and is optimized to capture semantic similarity. Beyond SBERT-FT there is a noticeable drop in performance with most other models achieving similar accuracy in the $\sim78\%-84\%$ range. More recent LLMS tend to perform better in this range which is expected as these models have a higher number of parameters. In addition we can observe the effect of finetuning, unsurprisingly finetuned versions outperform the pre-trained versions of any given model. 

In a different experiment, we used the same methodology with the GICS industry label instead of the sector as the target variable. The results of GICS industry prediction are also summarized in Table \ref{table:classification_sector_industry}. Even though this is a much harder classification task relative to sector prediction (we have 66 output classes instead of 11), the best model SBERT-FT-512 still achieves an F1 score of 0.79 which is notable. Similar to the sector case we see that SBERT-FT models perform the best while LF-PT models perform the worst. Examining the models that fall in the mid-range between these two extremes we see that finetuning seems to have a more prominent effect as opposed to model size as evidenced by the performance of LF-FT models. This can also be explained by the fact that the embedding models were finetuned at the industry level which was more granular compared to sector. Overall, this experiment validates that the embeddings generated by language models are semantically meaningful capturing the essential information needed to predict the sector or industry that a given company belongs to.

\textbf{Company Embeddings Outperform GICS Classification on downstream financial tasks.} Table \ref{table:corr} summarizes the performance of different models on similarity and return attribution tasks. While GICS industry/sector classification allows for bucketing companies in broad categories, company embeddings allow for a continuous distance metric to be calculated in embedding space, thus finding the closest peers of a given company. Particularly, the average pairwise correlation between companies in the same industry is 41\%. This is increased to 44\% and 47\% when looking at the top 5 and top 1 neighbors respectively. 


Additionally, as indicated by the $R^2$ column in table \ref{table:corr}, clusters of similar securities based on company  embeddings generated by the top-performing model can explain 11.9\% of cross-sectional monthly returns on average compared to 5.2\% and 10.6\% explained by GICS sector and industry respectively (i.e. ~12\% percentage improvement compared to traditional industry factors). Both experiments validate the ability of language models to capture the information discussed in the business description section effectively and generate financially meaningful embeddings which are similar for companies whose performance is impacted by similar market dynamics.

\textbf{Effect of pre-training objective.} BERT and LongFormer architectures lag behind SBERT and large language models in all 3 downstream tasks we studied. This can be explained by the fact that BERT and Longformer are optimized for text classification tasks and not for generating sentence embeddings. Specifically, BERT has been massively pretrained on classification objectives (masked language and next sentence prediction) while SBERT has been finetuned on semantic similarity tasks using contrastive loss ojectives which encourages the model to pull together embeddings of similar sentences and push apart embeddings of dissimilar ones. Thus, SBERT is more effective in generating high-quality sentence/document embeddings which are useful for downstream tasks that rely on semantic similarity, like the ones we explore in this paper. 

\textbf{Effect of model finetuning.} Additionally, this study aims to explore the effect of finetuning on the quality of the embeddings generated by a language model based on downstream tasks. As it shown in table \ref{table:corr}, finetuned models underperform compared to the pre-trained models on the similarity and return attribution tasks, while they outperformed pre-trained models on sector/industry classification tasks. Finetuning using GICS labels resulted in better embeddings only when the downstream task was aligned with the objective on which the model was finetuned, in this case predicting GICS sector/industry label. Instead, pre-trained model embeddings proved more effective for other downstream tasks which are not directly related to predicting GICS labels, since they capture the raw information in the business description without being biased in any way by GICS labels. 

\textbf{Effect of model size.} In this paper, we also aim to study the relationship between model size and the quality of the output embeddings. To this end, we compare the performance of the embeddings generated by SOTA large language models ($\sim 200-500$ billion parameters) with these generated by smaller models like SBERT ($\sim{400}$ million parameters). As shown in table \ref{table:corr}, PaLM and GPT marginally outperform SBERT on the return attribution task while they are on par on slightly worse than SBERT on the similarity task. Additionally, SBERT allows for finetuning which proves beneficial on sector/industry classification task outperforming PaLM and GPT. Consequently, while large language models set new standards on text generation tasks like question answering and summarization, significantly smaller models like SBERT can still generate equally meaningful text embeddings for financial applications as well as the flexibility to be finetuned on targeted tasks. 

\textbf{Effect of context window.} Overall experiments show that performance marginally increases for most models as the input context grows from 512 to 1536 tokens. For pre-trained models (SBERT, PaLM, GPT) accuracy is increased on the industry classification task since more granular information is provided to the model leading to richer embeddings, however, SBERT-FT achieved the greatest performance for input context of 512 tokens. This could be due to the fact that the model was finetuned based on input text of up to 512 tokens only. Similarly, on return attribution and similarity tasks, performance was slightly increased across models for larger context (for this reason we only report performance for context of 1536 tokens in table \ref{table:corr}), thus we conclude that more information can further enrich embeddings for similarity tasks.

\begin{table}[htbp]
\centering
\small
\begin{tabular}{l l l l  l l l}
\toprule
 & \multicolumn{3}{c}{Sector} & \multicolumn{3}{c}{Industry} \\

 & Acc. & \multicolumn{2}{c}{F1-Score} & Acc. & \multicolumn{2}{c}{F1-Score}\\
 &  & Micro & Weigh. &  & Micro & Weigh. \\
\midrule
SBERT-FT-1024   & \textbf{0.902} & \textbf{0.902} & \textbf{0.900} & 0.786 & 0.786 & 0.766 \\
SBERT-FT-1536   & 0.900 & 0.900 & 0.899 & 0.770 & 0.770 & 0.748 \\
SBERT-FT-512    & 0.900 & 0.900 & 0.899 & \textbf{0.793} & \textbf{0.793} & \textbf{0.773} \\
SBERT-1536      & 0.836 & 0.836 & 0.834 & 0.674 & 0.674 & 0.653 \\
PALM-1536       & 0.822 & 0.822 & 0.818 & 0.670 & 0.670 & 0.635 \\
GPT-1024        & 0.817 & 0.817 & 0.813 & 0.649 & 0.649 & 0.616 \\
GPT-512         & 0.815 & 0.815 & 0.811 & 0.637 & 0.637 & 0.603 \\
PALM-512        & 0.813 & 0.813 & 0.809 & 0.625 & 0.625 & 0.591 \\
LF-FT-1024      & 0.813 & 0.813 & 0.811 & 0.685 & 0.685 & 0.655 \\
GPT-1536        & 0.811 & 0.811 & 0.805 & 0.658 & 0.658 & 0.623 \\
PALM-1024      & 0.811 & 0.811 & 0.807 & 0.637 & 0.637 & 0.600 \\
SBERT-1024      & 0.811 & 0.811 & 0.808 & 0.681 & 0.681 & 0.665 \\
SBERT-512       & 0.809 & 0.809 & 0.805 & 0.662 & 0.662 & 0.639 \\
LF-FT-512       & 0.809 & 0.809 & 0.807 & 0.680 & 0.680 & 0.656 \\
BERT-FT-512     & 0.809 & 0.809 & 0.808 & 0.627 & 0.627 & 0.614 \\
BERT-FT-1536    & 0.805 & 0.805 & 0.804 & 0.633 & 0.633 & 0.625 \\
LF-FT-1536      & 0.803 & 0.803 & 0.799 & 0.687 & 0.687 & 0.659 \\
BERT-FT-1024    & 0.784 & 0.784 & 0.784 & 0.647 & 0.647 & 0.635 \\
LF-PT-1024      & 0.730 & 0.730 & 0.718 & 0.427 & 0.427 & 0.352 \\
LF-PT-512       & 0.726 & 0.726 & 0.713 & 0.438 & 0.438 & 0.361 \\
LF-PT-1536      & 0.716 & 0.716 & 0.704 & 0.432 & 0.432 & 0.359 \\

\bottomrule
\end{tabular}
\caption{Performance metrics for GICS Sector and Industry classification tasks.}
\label{table:classification_sector_industry}
\vspace{-8mm}
\end{table}

\section{Use Cases} \label{sec:usecases}
An important limitation of traditional industry classification schemes like GICS is their tendency to pigeonhole companies into a rigid, single ontology. This means that companies can only be assigned a single category at each level of the GICS categorization scheme. However, companies are multifaceted entities with operations that may span multiple sectors and industries. Despite its limitations, traditional industry classifications like GICS provide a well-established standard that is widely used and understood within the financial industry. Given its extensive application in numerous investment strategies and models, instead of shifting entirely to a new categorization scheme, an interesting approach to address the above limitation is to convert 'hard' assignments of GICS to 'soft' classifications. This can be achieved by using the embedding vectors as input features to predict GICS categories via supervised classification as described in section \ref{sec:sector_industry_classification}. However, instead of using the  model prediction, we take the class probabilities generated by the model. Using this approach instead of a single label, we obtain a probability distribution across each sector or industry for each company, which offers a more nuanced depiction of a company’s business.

\begin{figure}[h]
\centering
\includegraphics[width=0.45\textwidth]{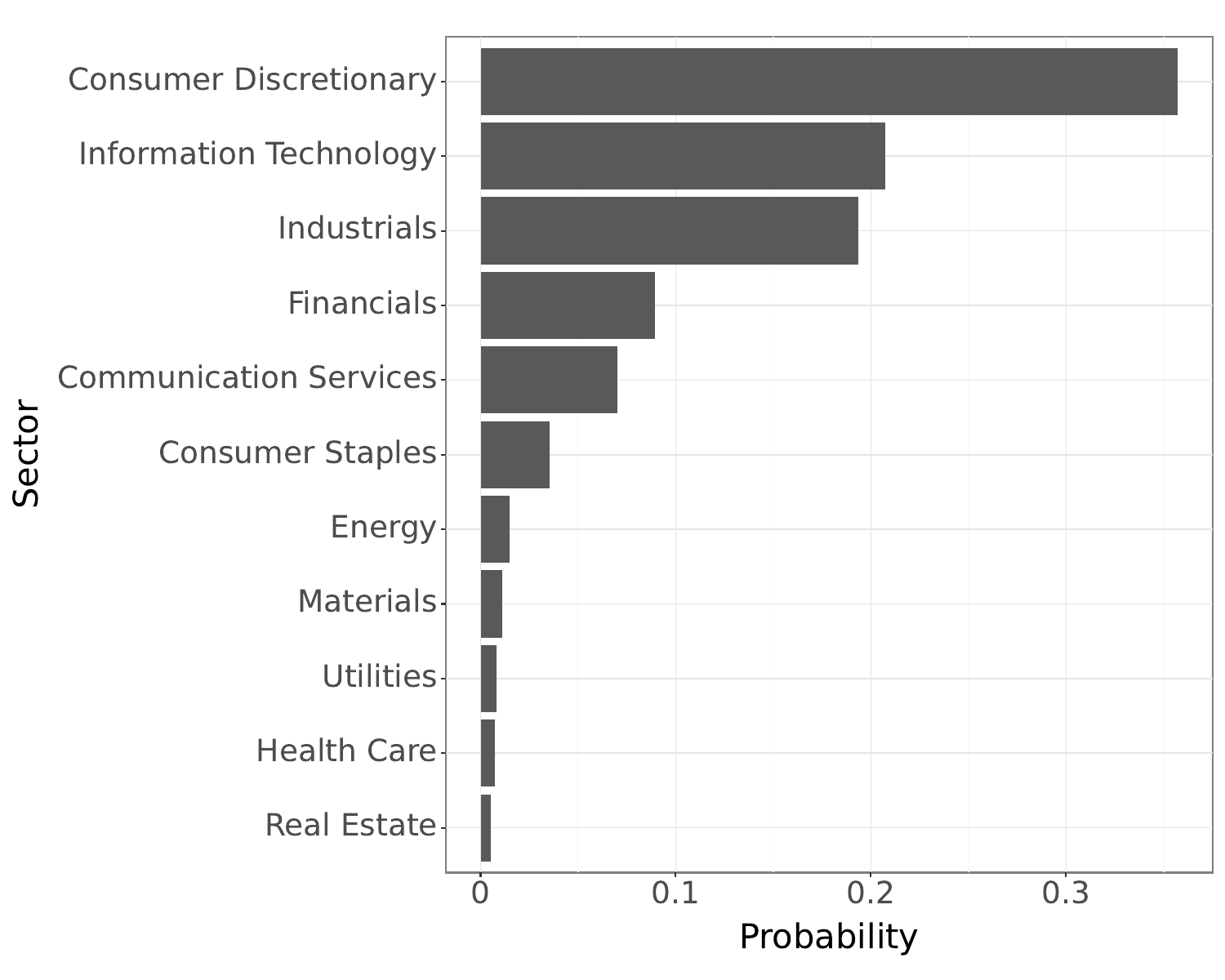}
\caption{Sector probabilities for Amazon}
\label{fig:amazon}
\end{figure}

Figure \ref{fig:amazon} illustrates the sector probabilities obtained for Amazon using the GPT-based embedding model using the first 1536 tokens as input text. We observe that the  three most likely sectors include Consumer Discretionary with a probability of 35.7\%, Information Technology at 20.7\%, and Industrials at 19.3\%. These results can be explained by Amazon's multifaceted business operations. Amazon's primary business sector is e-commerce, which fits within the Consumer Discretionary sector. As a part of its e-commerce business, Amazon also operates an extensive delivery and logistics infrastructure, which falls into the into the Industrials sector. Lastly, Amazon has been leading player in the cloud services industry with its Amazon Web Services (AWS) offering - which explains why the Information Technology sector appears with a high probability. 

\begin{figure}[h]
\centering
\includegraphics[width=0.45\textwidth]{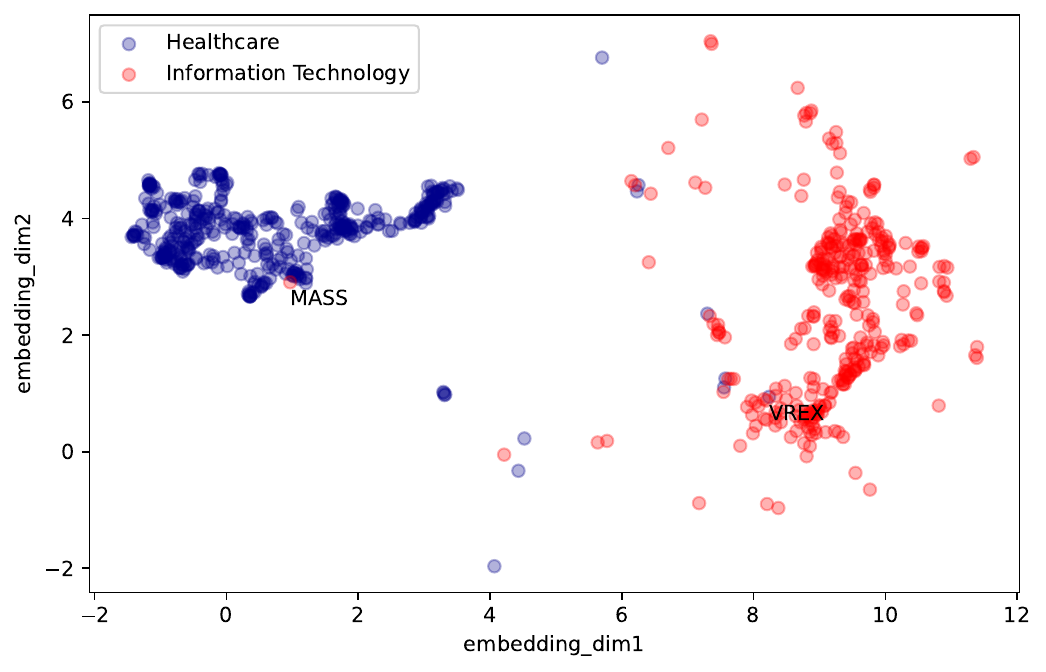}
\caption{Visualization of SBERT company embeddings after UMAP projection to 2D space.}
\label{fig:umap_plot}
\end{figure}

Additionally, embeddings are dense numerical representations which allow for continuous distance metrics to be calculated between different entities. Consequently, company embeddings can be used for outlier detection - companies which are further apart in the embedding space from the core of their sector assigned. To illustrate this idea, in figure \ref{fig:umap_plot} we plot SBERT embeddings for companies belonging to Information Technology and Healthcare. As an example of an outlier, MASS (908 Devices Inc) is a company assigned to Information Technology by GICS, however its embedding is much closer to the Healthcare cluster core. This can be explained due to the fact that MASS develops devices and technology for biopharmaceutical and healthcare applications. Similarly, VREX (Varex Imaging), which develops software and hardware for medical imaging applications, lies much closer to the Information Technology core in embedding space despite being assigned to Healthcare by GICS. While we do not explore outlier detection methods in this paper, an interesting future direction would be to exploit embeddings for detecting outlier companies and in turn use this information for better portfolio construction or idiosyncratic risk attribution.

\section{Conclusion}\label{sec:conclusion}
In this paper, we generate company embeddings using the raw business descriptions reported in SEC 10K filings using SOTA language models. We then benchmark the quality of these embeddings on multiple downstream financial tasks. 

First, we showed that we can reproduce GICS sector/industry classifications with high accuracy using the embeddings as features. Particularly, we demonstrated that we can optimize the quality of the embeddings by supervising the language models using sample company descriptions and the corresponding GICS industry labels, demonstrating that a small finetuned model can outperform modern pre-trained LLMs on this task. 

Secondly, we demonstrated that the company embeddings are financially meaningful since they be can used to find pairs or clusters of similar securities with high correlation in their daily returns. Specifically, using the embeddings we can identify the closest peer of each company in the embedding space which has 10\% greater returns correlation compared to the sector average and 6\% greater returns correlation compared to the industry average. Additionally, using embeddings we can form  clusters of similar securities which react in a similar way to systematic risk factors, thus explaining a larger percentage of cross-sectional equity returns compared to GICS sector and industry.

A limitation of company embeddings generated by language models compared to traditional sector/industry classifications is the lack of interpretability. However, company embeddings generated based on textual information can be a valuable tool for downstream financial applications. For example, GICS classifications are only available for public companies. Instead, company embeddings can be generated based on the business description of a private company too, thus providing a way to perform similarity learning in private markets or finding the closest private/public peers of any private company. Additionally, company embeddings can be used as numerical features in downstream models for soft industry classification or outlier detection. Finally, one interesting direction for future work would be to explore supervised company similarity models using both tabular data as well as the embeddings extracted from text data as input features.

\section*{Acknowledgement}
The views expressed here are those of the authors alone and not of BlackRock, Inc.

\bibliographystyle{ACM-Reference-Format}
\bibliography{sample-base}
\end{document}